# Method for estimating modulation transfer function from sample images

Rino Saiga [a], Akihisa Takeuchi [b], Kentaro Uesugi [b], Yasuko Terada [b], Yoshio Suzuki [c], and Ryuta Mizutani [a,*]

[a] Department of Applied Biochemistry, School of Engineering, Tokai University, Kitakaname 4-1-1, Hiratsuka, Kanagawa 259-1292, Japan
[b] Japan Synchrotron Radiation Research Institute (JASRI/SPring-8), Kouto 1-1-1, Sayo, Hyogo 679-5198, Japan
[c] Graduate School of Frontier Sciences, The University of Tokyo, Kashiwanoha 5-1-5, Kashiwa, Chiba 277-8561, Japan
**Corresponding author*: Ryuta Mizutani, Department of Applied Biochemistry, Tokai University, Hiratsuka, Kanagawa 259-1292, Japan. Tel: +81-463-58-1211; Fax: +81-463-50-2426; e-mail: ryuta@tokai-u.jp

**Abstract:** The modulation transfer function (MTF) represents the frequency domain response of imaging modalities. Here, we report a method for estimating the MTF from sample images. Test images were generated from a number of images, including those taken with an electron microscope and with an observation satellite. These original images were convolved with point spread functions (PSFs) including those of circular apertures. The resultant test images were subjected to a Fourier transformation. The logarithm of the squared norm of the Fourier transform was plotted against the squared distance from the origin. Linear correlations were observed in the logarithmic plots, indicating that the PSF of the test images can be approximated with a Gaussian. The MTF was then calculated from the Gaussian-approximated PSF. The obtained MTF closely coincided with the MTF predicted from the original PSF. The MTF of an x-ray microtomographic section of a fly brain was also estimated with this method. The obtained MTF showed good agreement with the MTF determined from an edge profile of an aluminum test object. We suggest that this approach is an alternative way of estimating the MTF, independently of the image type.

*Abbreviations:* CCD, charge-coupled device; FWHM, full width at half maximum; LSF, line spread function; MTF, modulation transfer function; PSF, point spread function

## 1. Introduction

The modulation transfer function (MTF) is a widely used measure of the frequency domain response of imaging modalities, including microscopy (Faruqi & McMullan, 2011), radiology (Rossmann, 1969; Workman & Brettle, 1997), and remote sensing (Rauchmiller & Schowengerdt, 1988). The MTF is defined as the norm of the Fourier transform of the point spread function (PSF). The PSF of an image represents blurring accompanying the visualization process of the image. Hence, spatial resolution can be evaluated from the high frequency profile of the MTF.

A number of methods for estimating the MTF have been reported. The primary method for estimating the MTF is to determine the PSF and then transform it into the MTF. The PSF has been measured using small objects, whose dimension is negligible relative to the spatial resolution (Rauchmiller & Schowengerdt, 1988). An alternative to the PSF is the line spread function (LSF) that can be determined from the edge profile (Judy, 1976; Mizutani et al., 2010a). Although these approaches provide methods to determine the MTF, a test object appropriate for the target resolution must be prepared for measuring the PSF or the LSF. Another concern is that the MTF of the test object image cannot exactly represent that of the sample image. This is because the image resolvability depends not only on the optical performance, but also on many other factors, such as apparatus vibration, temperature fluctuation, and sample deformation.

An MTF estimation without using test objects has been reported for a satellite imaging system (Delvit et al., 2004). The MTF of the satellite camera was estimated with an artificial neural network, which learned the association between simulated images and their MTFs. The MTF has also been estimated by using fractal characteristics of natural images (Xie et al., 2015). Although the MTFs of general images can be estimated with these methods, the theoretical relationships between the image characteristics and the obtained MTFs have not been fully delineated. Since the image characteristics depend on the sample or the target object, the feasibility of knowledge-based methods should be examined for each imaging modality or each image type. The presampling two-dimensional MTF was determined from the noise power spectrum of the detector (Kuhls-Gilcrist et al., 2010), though multiple flat-field images had to be taken at several different exposure levels.

In this study, we report a method for estimating the MTF from general sample images independently of the image type. It has been reported that the Gaussian PSF can be identified by plotting the logarithm of the squared norm of the Fourier transform of the image against the squared distance from the origin (Mizutani et al., 2016). A number of test images were generated and analyzed with this plot to extract their PSFs. The MTF was then calculated from the PSF. The obtained results indicated that the MTF can be estimated from the image itself. The MTF of an x-ray microtomographic section of a fly brain was evaluated with this method and compared with the MTF determined from an edge profile and with a tomographic cross section of square-wave test patterns.

## 2. Materials and Methods

### 2.1 Test images

Ultrathin sections of an epoxy block of a mouse brain fixed with formaldehyde, glutaraldehyde and osmium tetroxide were stained with uranyl acetate and Sato's lead reagent (Sato, 1968). Section images were taken with a transmission electron microscope (JEM-1200EX, JEOL) operated at 80 kV. The obtained image was digitized and subjected to 4 × 4 binning in order to eliminate intrinsic blurring due to the microscope optics. The resultant image dimensions were 512 × 512 pixels.

A paraffin section of a zebrafish brain stained with the reduced-silver impregnation (Mizutani et al., 2008) was taken with a light microscope (Eclipse80i, Nikon) equipped with a charge-coupled device (CCD) camera (DXM1200F, Nikon), as reported previously (Mizutani et al., 2016), and subjected to 4 × 4 binning. The image dimensions were 500 × 500 pixels.

A satellite image of Tokyo Bay was clipped from an Earth observation image (P-012-14249, JAXA Digital Archives) taken by the ALOS observation satellite (Shimada et al., 2010), and subjected to 2 × 2 binning. The image dimensions were 252 × 252 pixels.

The color images were converted into grayscale ones prior to the binning. Test images were generated by convolving these original images and circular aperture PSFs or a Gaussian PSF. Before the PSF convolution, each pixel was subdivided into 4 × 4 subpixels in order to reproduce the PSF shape in the digitized image. The PSF-convolved images were then re-scaled to the original dimensions and subjected to a Fourier transformation. In order to minimize truncation errors in the Fourier transforms, image edges were smoothed to the average of the pixel intensities and margin pixels were filled with the average intensity.

### 2.2 MTF estimation from sample images

We have reported a method for estimating the full width at half maximum (FWHM) of the PSF from sample images (Mizutani et al., 2016). By assuming a Gaussian PSF, the width of the PSF can be determined from a relationship:

$$\ln\left|F(\mathbf{k})\right|^2 \cong -4\pi^2\sigma^2\left|\mathbf{k}\right|^2 + \text{constant}\ ,$$

where $F(\mathbf{k})$ is the Fourier transform of the image and σ is the standard deviation of the Gaussian PSF. This equation indicates that the Gaussian PSF can be identified as a linear correlation by plotting $\ln|F(\mathbf{k})|^2$ as a function of $|\mathbf{k}|^2$. The logarithm of the squared norm of the Fourier transform of the image was calculated using the resolution plot function of the RecView program (Mizutani et al., 2010a; available from https://mizutanilab.github.io/). The obtained PSF was then transformed into the MTF.

### 2.3 Samples for x-ray tomographic microscopy

An aluminum test object with square-wave patterns was prepared by using a focused ion-beam (FIB) apparatus (FB-2000, Hitachi High-Technologies), as reported previously (Mizutani et al., 2010b). An aluminum wire with an approximate diameter of 250 μm was subjected to ion-beam milling. A gallium beam of 15 nA was used for rough abrasion of the aluminum surface, and a 2 nA beam was used for finishing the flat face. A series of square wells was carved on the flat face. The pitches of the square-wave patterns were 2.0, 1.6, 1.2, 1.0, 0.8, and 0.6 μm. Each pattern was composed of half-pitch wells and half-pitch intervals, *i.e.*, a 0.5 μm well and a 0.5 μm interval for a 1.0 μm pitch. The aluminum wire was recovered and its flat face was coated with Petropoxy 154 epoxy resin (Burnham Petrographics, ID) to avoid the mechanical damage of the patterns.

A fly brain sample was prepared from wild-type fruit fly *Drosophila melanogaster* Canton-S (Drosophila Genetic Resource Center, Kyoto Institute of Technology). Adult fly brains were dissected and subjected to the reduced-silver impregnation, as reported previously (Mizutani et al., 2013). The sample was dehydrated in an ethanol series, then transferred to n-butyl glycidyl ether, and finally equilibrated with Petropoxy 154 epoxy resin. The obtained sample was mounted on a nylon loop (Hampton Research, CA).

### 2.4 X-ray tomographic microscopy

X-ray tomographic microscopy was performed at the BL47XU beamline of SPring-8, as described previously (Takeuchi et al., 2006). The sample was mounted on the rotation stage by using a brass fitting specially designed for biological samples. A Fresnel zone plate with outermost zone width of 100 nm and diameter of 774 μm was used as an x-ray objective lens. Transmission images produced by 8-keV x-rays were recorded using a CCD-based imaging detector (AA40P and C4880-41S, Hamamatsu Photonics, Japan). The viewing field was 2000 pixels horizontally × 700 pixels vertically. The effective pixel size was 262 nm × 262 nm. One dataset consisted of 3000 image frames. Each frame was acquired with a rotation step of 0.06º and exposure time of 300 ms. The obtained x-ray images were subjected to a convolution-back-projection calculation with the RecView program (Mizutani et al., 2010a) to reconstruct tomographic cross sections.

## 3. Results

### 3.1 MTF estimation using test images

An electron microscopy image of a mouse brain section, in which myelins were observed as ring-like structures (Fig. 1a), was used for generating test images. A circular aperture PSF having a radius of 2 pixels or 4 pixels, or a Gaussian PSF having an FWHM of 6 pixels was convolved with this original image. The resultant images (Fig. 1c, 1e, and 1g) showed blurring due to the PSFs. Fourier transforms of these test images are shown in Figures 1b, 1d, 1f, and 1h. The Fourier transform of the original image showed a monotonically decreasing profile (Fig. 1b), while those of the convolved images exhibited Airy disks or a simple peak depending on the applied PSFs (Fig. 1d, 1f, and 1h).

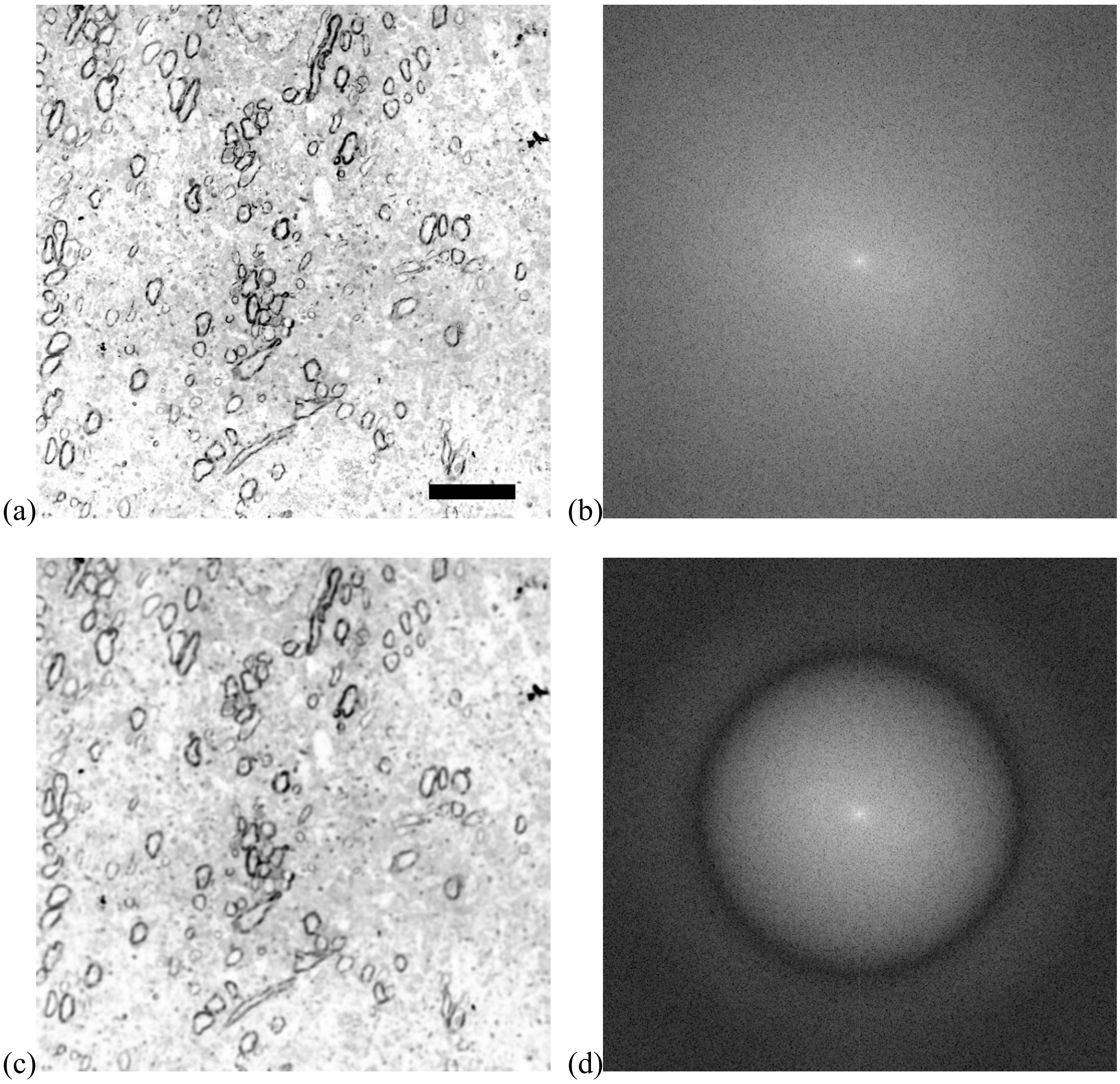

**Figure 1.** (**a**) Electron microscopy image of a mouse brain section. Scale bar: 5 μm. (**b**) Fourier transform of the brain section image. The distribution of the norm of the Fourier transform is illustrated in a logarithmic gray scale. (**c**) Test image generated by convolving the brain section image shown in panel **a** and a circular aperture PSF having a radius of 2 pixels. (**d**) Fourier transform of the test image shown in panel **c**.

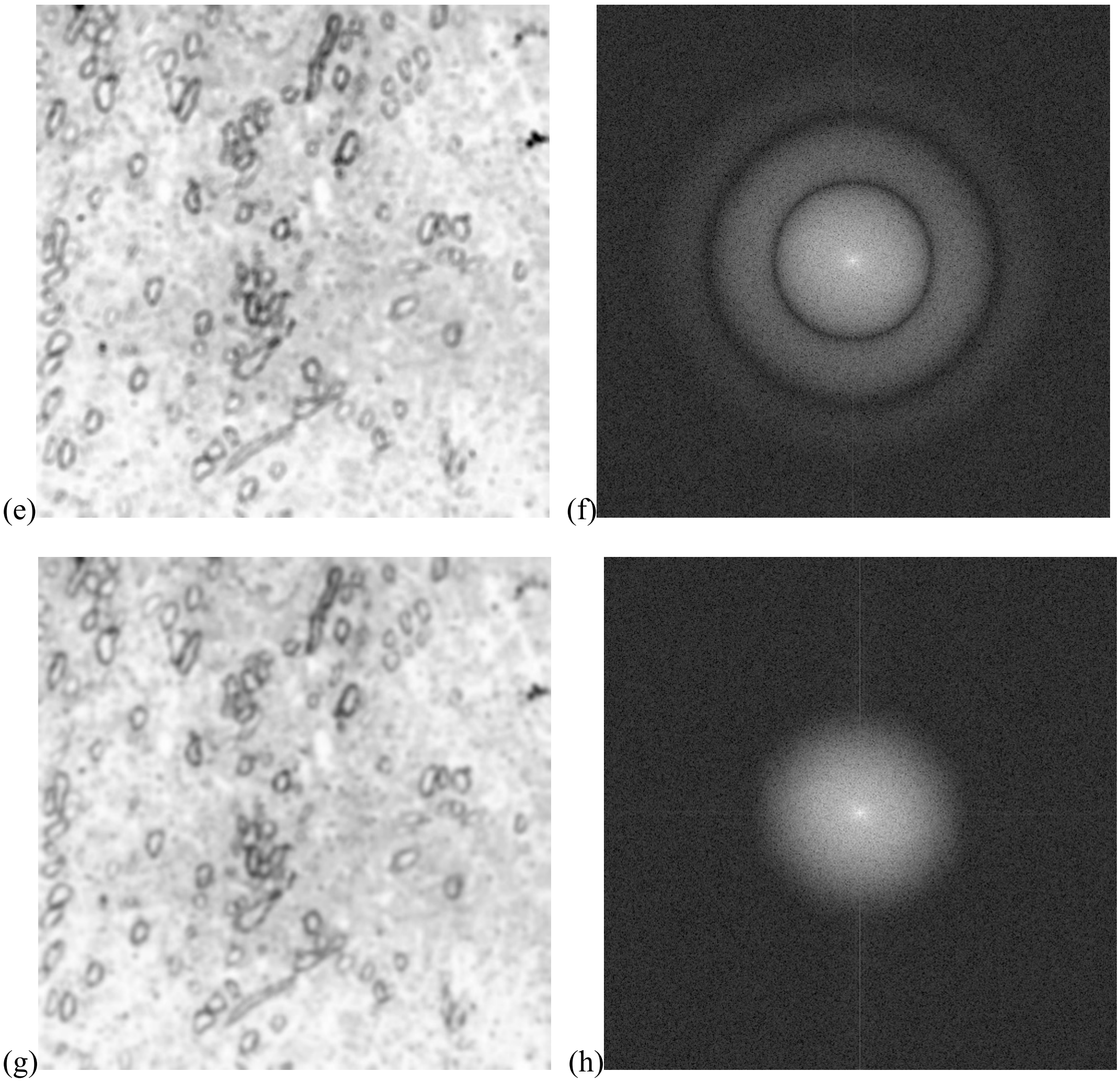

**Figure 1 cont'd.** (**e**) Test image generated by using a circular aperture PSF with a radius of 4 pixels. (**f**) Fourier transform of the test image shown in panel **e**. (**g**) Test image generated by using a Gaussian PSF with an FWHM of 6 pixels. (**h**) Fourier transform of the test image shown in panel **g**. Streaks along the vertical or horizontal axes are due to image truncation.

The logarithm of the squared norm of these Fourier transforms was plotted against the distance from the origin in the frequency domain (Fig. 2a, 2c, and 2e). We have reported that Gaussian-approximated PSFs can be extracted from sample images using the logarithmic plot (Mizutani et al., 2016). The obtained plots showed profiles corresponding to the circular aperture PSF or the Gaussian PSF. The left ends of the plots can be regarded to have linear correlations (Fig. 2a, 2c, and 2e). This suggests that the PSFs of the test images can be approximated with Gaussians. The region for evaluating the linear correlation coefficient was defined to be 80% of the first lobe of the Airy disk (Fig. 2a and 2c), or it was defined from a kink between the linear correlation and the noise floor (Fig. 2e). MTFs estimated from the Gaussian-approximated PSFs are shown in Figure 2b, 2d, and 2f. The estimated MTFs are in good agreement with the MTFs predicted from the original PSFs, though the MTF estimated from the test image of the 2-pixel radius aperture was lower than the predicted MTF. The maximum difference between the estimated and predicted MTFs was 0.083. The difference is ascribable to an error due to approximating the PSF with a Gaussian using the linear correlation in the plot.

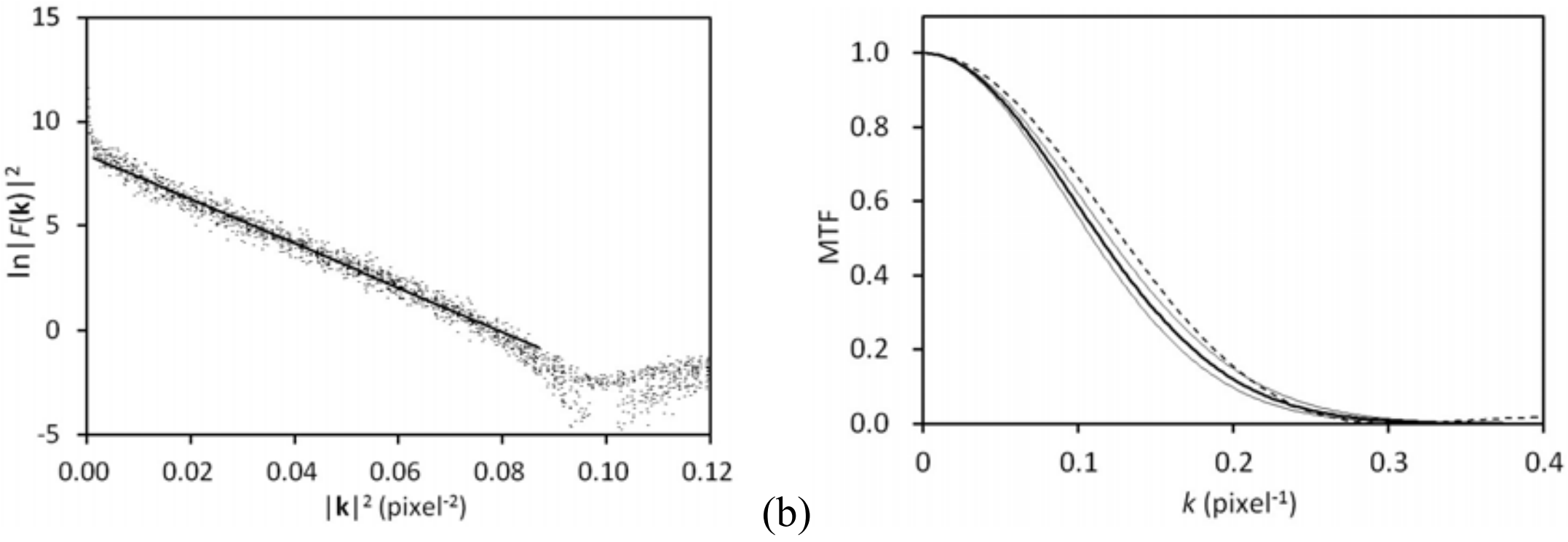

(a) (b)

**Figure 2.** (**a**) Radial profile of the Fourier transform of the mouse brain section convolved with a circular aperture PSF having a radius of 2 pixels. The logarithm of the squared norm of the Fourier transform is plotted against the squared distance from the origin. Each 5 × 5 pixel bin was averaged to reduce data points. The linear correlation indicated with the solid line represents a Gaussian approximation of the PSF. (**b**) MTF calculated from the Gaussian PSF estimated in panel **a**. The thick line indicates the MTF calculated from the Gaussian. Thin lines correspond to ±10% deviations of the linear correlation coefficient in the Gaussian approximation. The dashed line represents the MTF predicted from the circular aperture PSF.

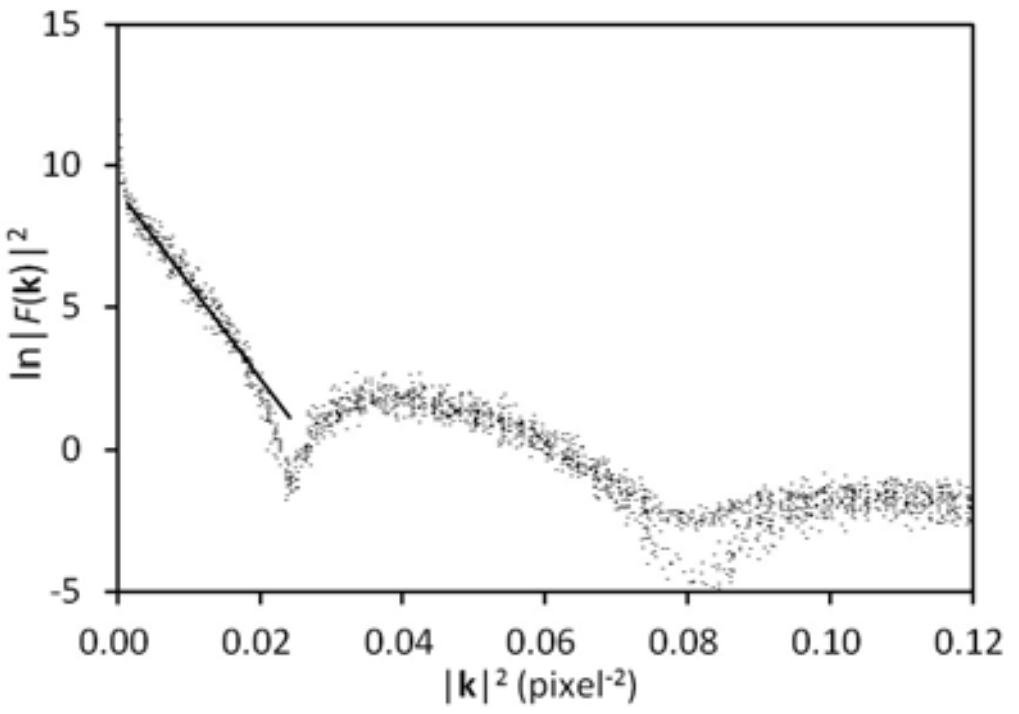

(c)

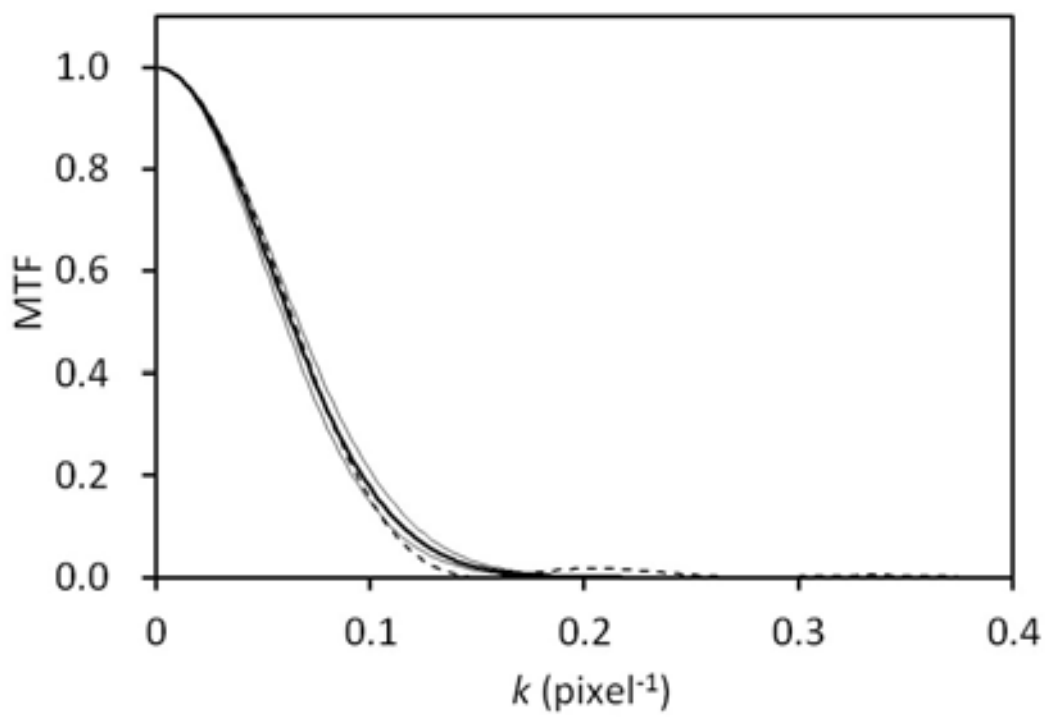

(d)

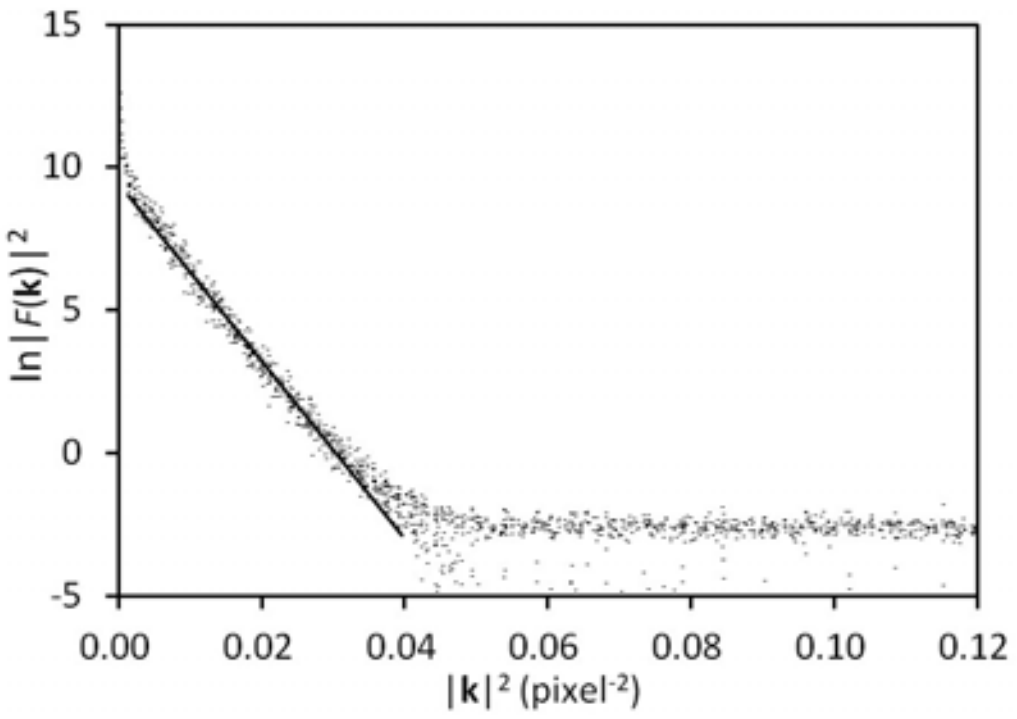

(e)

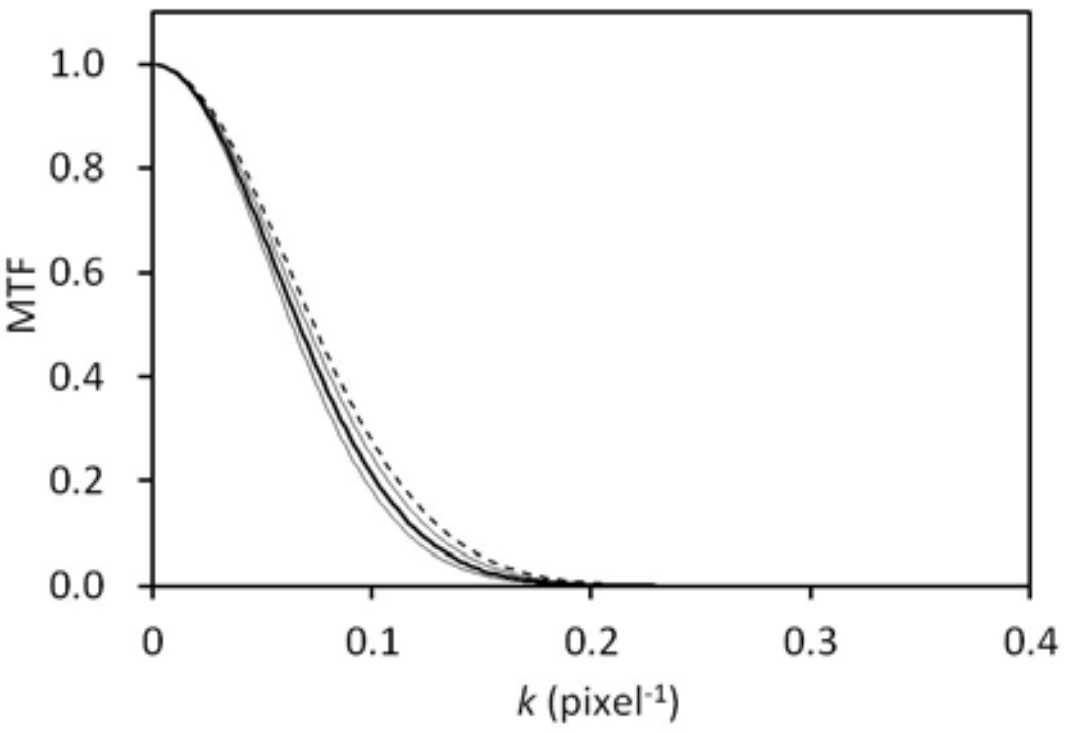

(f)

**Figure 2 cont'd.** (**c**) Radial profile of the Fourier transform of the brain section image convolved with a circular aperture PSF having a radius of 4 pixels. (**d**) MTF calculated from the Gaussian PSF estimated in panel **c**. (**e**) Radial profile of the Fourier transform of the brain section image convolved with a Gaussian PSF having an FWHM of 6 pixels. (**f**) MTF calculated from the Gaussian PSF estimated in panel **e**.

Figure 3 shows the application of this method to a number of image types. A paraffin section of a zebrafish brain (Fig. 3a) and a satellite image of Tokyo Bay (Fig. 3c) were convolved with a circular aperture PSF having a radius of 2 pixels. MTFs of these test images were estimated from the logarithmic plots of their Fourier transforms (Figs. 3b, 3d). Although the MTFs were slightly underestimated, the obtained plots nearly reproduced the MTF predicted from the PSF.

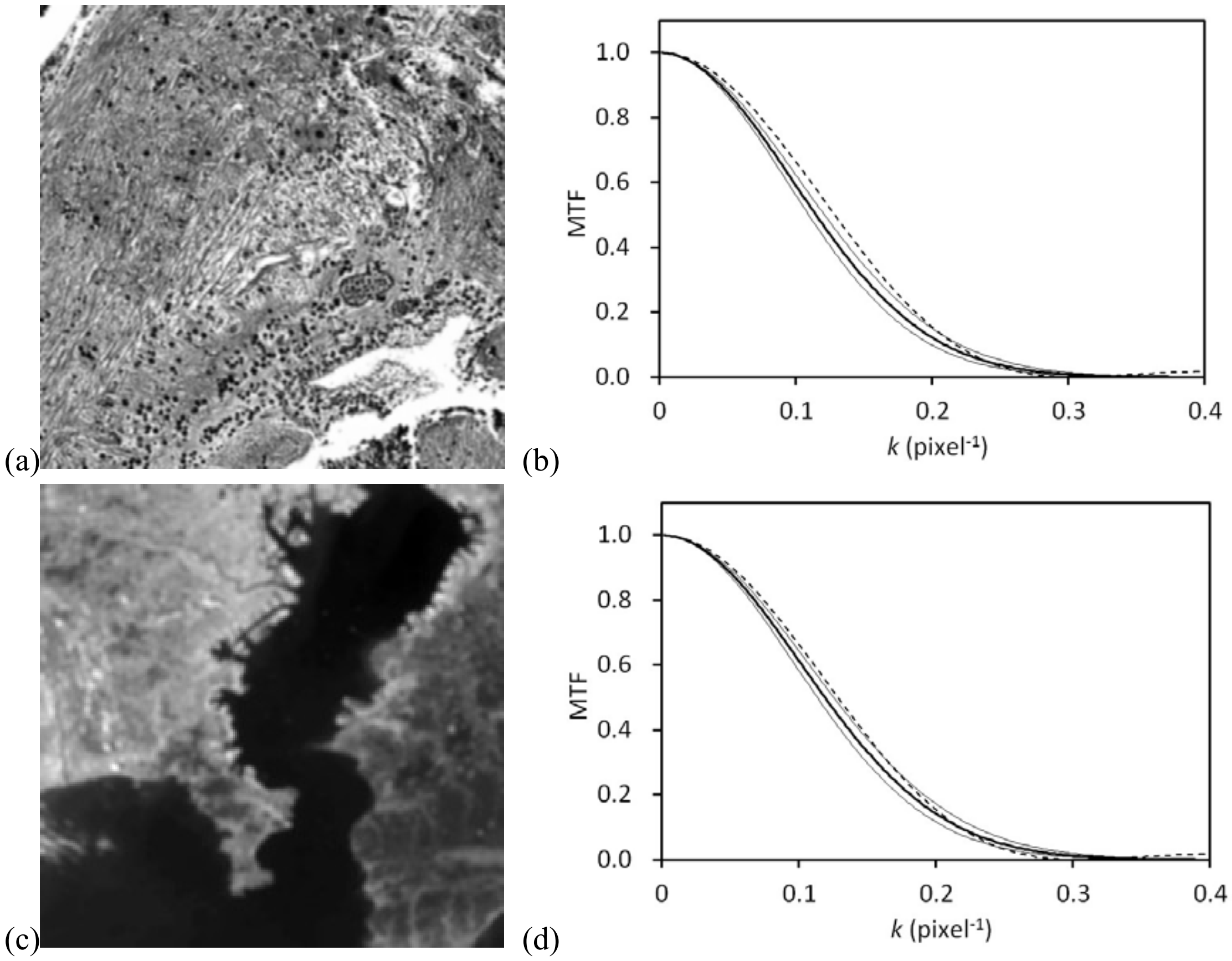

**Figure 3.** Test image of a paraffin section of a zebrafish brain (**a**) and satellite image of Tokyo Bay (**c**) all generated by convolving original images and a circular aperture PSF having a radius of 2 pixels. The MTFs of these images (**b** and **d**, respectively) were estimated with the logarithmic plot by approximating the PSF with Gaussians. Thick lines indicate MTFs calculated from the Gaussian-approximated PSFs. Thin lines correspond to ±10% deviations of the linear correlation coefficients in the Gaussian approximation. Dashed lines represent the MTF predicted from the circular aperture PSF.

### 3.2 MTF estimation using real sample images

The resolution of an x-ray microtomography system equipped with Fresnel zone plate optics was evaluated using square-wave patterns. Figure 4a shows a cross section of an aluminum test object visualized with x-ray microtomography. The square wave patterns up to a pitch of 1.0 μm were resolved. The MTF of the tomographic cross section was also determined with the slanted edge method (Judy, 1976) by using a flat face of the aluminum test object. Figure 4b shows an edge profile along with LSF calculated from the profile. The FWHM of the obtained LSF was 1.0 μm, which coincides with the resolution of the square-wave patterns.

Figure 4c shows a tomographic cross section of a brain of a fruit fly *Drosophila melanogaster* visualized with the same apparatus. Neuronal processes in the middle of the optic lobe along with spherical somas surrounding them were observed in this section. The logarithm of the squared norm of the Fourier transform of this image was plotted to extract the PSF. Figure 4d shows the logarithmic plot for the fly brain section. The PSF of a real sample image depends on a number of factors including the optical performance, apparatus vibration, temperature fluctuation, and sample deformation. Although it is difficult to formulate all of these factors, a Gaussian function is a practical approximation for representing the actual PSF. A linear correlation was observed at the left end of the logarithmic plot, indicating that the PSF of this image can be indeed approximated with a Gaussian. The MTF was then calculated from the Gaussian PSF.

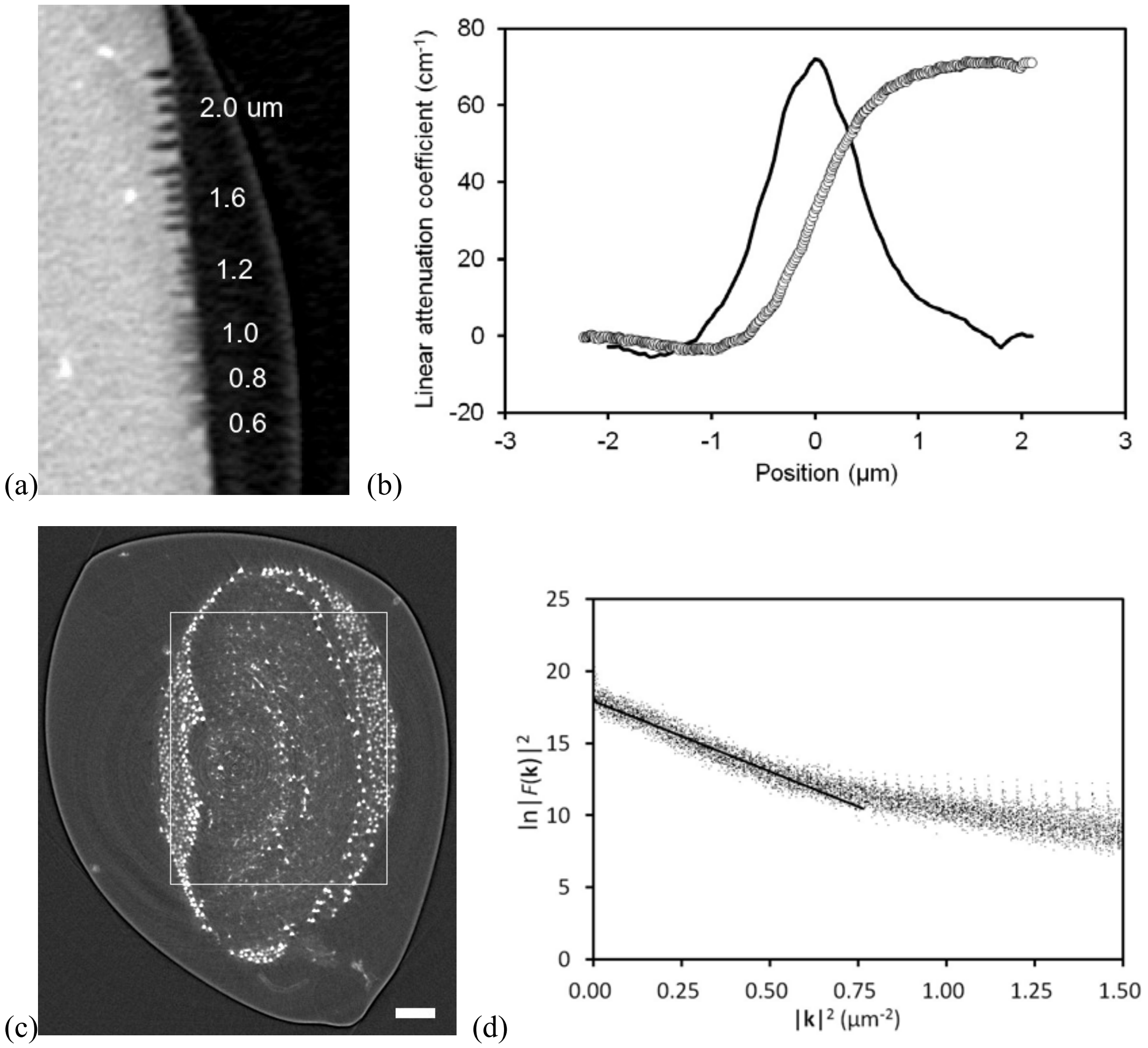

**Figure 4.** (**a**) Tomographic cross section of square-wave patterns with pitches of 2.0, 1.6, 1.2, 1.0, 0.8, and 0.6 μm carved on a flat face of an aluminum wire. Patterns up to 1.0 μm pitch were resolved. (**b**) Edge profile of the flat face of the aluminum test object. Open circles indicate linear attenuation coefficients of image pixels plotted against the distance from the edge surface. The solid line indicates the LSF calculated from the edge profile. (**c**) Tomographic cross section of a fly brain embedded in epoxy resin. The white box indicates the region used for the MTF estimation. Scale bar: 20 μm. (**d**) Logarithmic plot of the squared norm of the Fourier transform of the boxed region in panel **c**. Each 5 × 5 pixel bin was averaged to reduce the number of data points. The linear correlation indicated with the solid line demonstrates that the PSF can be approximated with a Gaussian. Spikes in the right half originated from the image truncation.

The MTF estimated from the logarithmic plot along with the MTF calculated from the edge profile are superposed in Figure 4e. These MTFs showed consistent profiles, though the MTF of the brain section image was slightly lower than the MTF of the edge profile. A possible cause of this difference is the underestimation associated with this method. In the test image evaluation (*e.g.*, Fig. 2b), the MTFs were estimated to be slightly lower than the predicted MTF. Therefore, the difference may be due to this tendency.

Another factor in the underestimation is sample drift or deformation. Since tomographic cross sections are reconstructed from a series of images, sample drift or deformation during the dataset acquisition can affect the MTF. One of the raw images used for the tomographic reconstruction of the fly brain section was subjected to the MTF estimation with the same procedure (Fig. 4e). The obtained MTF of the raw image performed better than those calculated from the tomographic cross section of the fly brain and from the edge profile of the test object. These results suggest that the MTFs of the tomographic cross sections suffered from sample drift or stage fluctuation during the dataset acquisition.

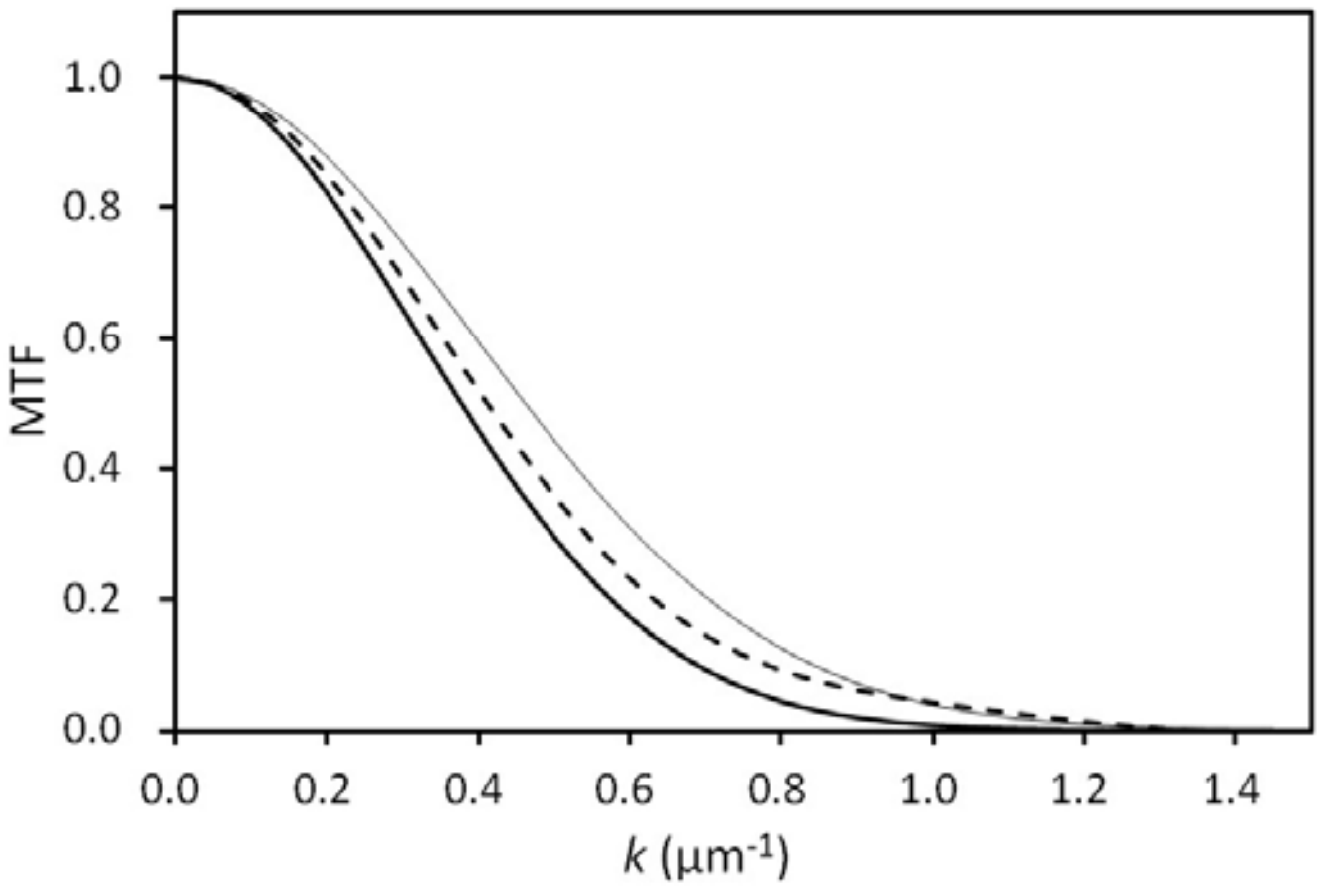

(e)

**Figure 4 cont'd.** (**e**) MTF calculated from the Gaussian-approximated PSF. The thick line indicates the MTF estimated from the boxed region in panel **c**. The dashed line represents the MTF calculated from the edge profile shown in panel **b**. The thin line represents the MTF estimated from one of the raw images used for the tomographic reconstruction of the image shown in panel **c**.

**4. Discussion**

Image resolution is the major concern of visualization methods. The MTF is a measure of image resolvability in the frequency domain (Rossmann, 1969). MTFs of imaging systems have been determined using objects appropriate for the MTF estimation (Judy, 1976; Cunningham & Reid, 1992; Fujita et al., 1992; Storey, 2001; Mizutani et al., 2010a). However, these methods use a small part of the image, such as an edge or a spot in the image. The MTF estimated from a small part of the image does not represent the MTF of the entire image, since the optical aberration can impair the resolution of other parts of the image. Moreover, the MTF of the real sample image depends not only on the optical performance, but also on many other factors related to ambient conditions and to the sample itself. Therefore, the MTF determined using a test object cannot be exactly the same as the MTF of the sample image.

It has been reported that an artificial neural network can be trained to recognize the MTF of nonspecific images (Delvit et al., 2004), though the neural network must be trained in advance with a set of images. The MTF has also been estimated by exploiting the fractal nature of landscape images (Xie et al., 2015), although this method depends on the specific character of the target images. In contrast, the approach reported in this paper allows the MTF to be estimated from general images without prior knowledge. The obtained results indicated that MTFs can be estimated from a number of test and real sample images.

The prerequisite of this method is that the MTF of the target image must be approximated with a Gaussian. If the linear correlation corresponding to the Gaussian is not identified in the logarithmic plot of the Fourier transform of the image, the PSF cannot be extracted, and hence, the MTF cannot be estimated from the sample image. Another limitation is the difficulty in estimating the MTF from sparse images. A sparse image results in a characteristic Fourier transform, in which the noise profile dominates the entire logarithmic plot (Mizutani et al., 2016). This makes it difficult to identify the linear correlation at the left end of the plot. In such cases, the sample region can be clipped to minimize the background area, so that the MTF can be estimated from the region of interest.

In a similar manner, the MTF of an arbitrary part of the image can be estimated by clipping that part and calculating its Fourier transform. Therefore, even in the case that the PSF locally changes within the image, the MTF of each part can be separately determined with this method. If the imaging system has astigmatism, its MTF depends on the direction in the image. In this study, we assumed that the PSF exhibits the same profile in every direction and plotted the logarithmic profile as a radial distribution. When astigmatism is observed in the image, multiple logarithmic plots along different directions can be generated so that we can estimate the MTF as a function of the direction.

The MTF estimated with this method showed a tendency to be slightly underestimated. This is presumably due to the origin spike of the Fourier transform. The origin of the Fourier transform is equal to the sum of pixel values of the entire image, while the outer periphery of the Fourier transform converges to the PSF. For example, an Airy disk was observed in the outer periphery of Figure 1d and 1f, whereas the origin exhibits a spike that does not appear in the logarithmic profile of the Airy disk. The origin spike should cause steepening of the linear correlation in the logarithmic plot, resulting in broadening of the Gaussian-approximated PSF and consequently underestimation of the MTF. These underestimations slightly varied depending on image types (Figs. 2 and 3). Since our method assumes that the sample has a random structure, the differences in the estimation errors are ascribable to differences in the degree of the structural randomness.

## 5. Conclusions

In order to estimate the MTF using a test object, the dimension or the precision of the test object must be much finer than the target resolution. The imaging system should also keep its conditions constant between the acquisitions of the sample and test object images. Even if such requirements are satisfied, deformation of the sample or its position in the viewing field would affect the image. Therefore, the MTF of a sample image is not exactly the same as the MTF estimated using a test object. In this study, we estimated the MTF by using the logarithmic plot of the Fourier transform of the image. The obtained results indicated that the MTF can be estimated from the sample image itself. We suggest that this approach is an alternative way of estimating the MTF, independently of the image type or imaging modality.

## Acknowledgements

We thank Yoshiko Itoh and Noboru Kawabe (Teaching and Research Support Center, Tokai University School of Medicine) for their helpful assistance with the electron microscopy and histology. This study was supported in part by Grants-in-Aid for Scientific Research from the Japan Society for the Promotion of Science (nos. 25282250 and 25610126). The synchrotron radiation experiments were performed at SPring-8 with the approval of the Japan Synchrotron Radiation Research Institute (JASRI) (proposal nos. 2008B1261 and 2017A1143).

## References

Cunningham, I.A., Reid, B.K., 1992. Signal and noise in modulation transfer function determinations using the slit, wire, and edge techniques. Med. Phys. 19, 1037-1044.

Delvit, J.-M. Leger, D., Roques, S., Valorge, C., 2004. Modulation transfer function estimation from nonspecific images. Opt. Eng. 43, 1355-1365.

Faruqi, A.R., McMullan, G., 2011. Electronic detectors for electron microscopy. Q. Rev. Biophys. 44, 357-390.

Fujita, H., Tsai, D.-Y., Itoh, T., Doi, K., Morishita, J., Ueda, K., Ohtsuka, A., 1992. A simple method for determining the modulation transfer function in digital radiography. IEEE Trans. Medical Imaging 11, 34-39.

Judy, P.F., 1976. The line spread function and modulation transfer function of a computed tomographic scanner. Med. Phys. 3, 233-236.
Kuhls-Gilcrist, A., Jain, A., Bednarek, D.R., Hoffmann, K.R., Rudin, S., 2010. Accurate MTF measurement in digital radiography using noise response. Med. Phys. 37, 724-735.
Mizutani, R., Takeuchi, A., Uesugi, K., Takekoshi, S., Osamura R.Y., Suzuki, Y., 2008. X-ray microtomographic imaging of three-dimensional structure of soft tissues. Tissue Eng. Part C 14, 359-363.
Mizutani, R., Taguchi, K., Takeuchi, A., Uesugi, K., Suzuki, Y., 2010a. Estimation of presampling modulation transfer function in synchrotron radiation microtomography. Nucl. Instrum. Meth. A 621, 615-619.
Mizutani, R., Takeuchi, A., Osamura, R.Y., Takekoshi, S., Uesugi, K., Suzuki, Y., 2010b. Submicrometer tomographic resolution examined using a micro-fabricated test object. Micron 41, 90-95.
Mizutani, R., Saiga, R., Takeuchi, A., Uesugi, K., Suzuki, Y., 2013. Three-dimensional network of Drosophila brain hemisphere. J. Struct. Biol. 184, 271-279.
Mizutani, R., Saiga, R., Takekoshi, S., Inomoto, C., Nakamura, N., Itokawa, M., Arai, M., Oshima, K., Takeuchi, A., Uesugi, K., Terada, Y., Suzuki, Y., 2016. A method for estimating spatial resolution of real image in the Fourier domain. J. Microsc. 261, 57-66.
Rauchmiller, R.F., Schowengerdt, R.A., 1988. Measurement of the landsat thematic mapper modulation transfer function using an array of point sources. Opt. Eng. 27, 274334.
Rossmann, K., 1969. Point spread-function, line spread-function, and modulation transfer function: Tools for the study of imaging systems. Radiology 93, 257-272.
Sato, T., 1968. A modified method for lead staining of thin sections. J. Electron. Microsc. 17, 158-159.
Shimada, M., Tadono, T., Rosenqvist, A., 2010. Advanced Land Observing Satellite (ALOS) and monitoring global environmental change. Proc. IEEE 98, 780-799.
Storey, J.C., 2001. Landsat 7 on-orbit modulation transfer function estimation. Proc. SPIE 4540, doi:10.1117/12.450647.
Takeuchi, A., Uesugi, K., Suzuki, Y., Tamura, S., Kamijo, N., 2006. High-resolution x-ray imaging microtomography with Fresnel zone plate optics at SPring-8. IPAP Conf. Series 7, 360-362.
Workman, A., Brettle, D.S., 1997. Physical performance measures of radiographic imaging systems. Dentomaxillofac. Radiol. 26, 139-146.
Xie, X., Wang, H., Zhang, W., 2015. Statistic estimation and validation of in-orbit modulation transfer function based on fractal characteristics of remote sensing images. Opt. Commun. 354, 202-208.

**Highlights**

We estimated the MTF from the sample image itself.

MTF of a micro-CT section closely coincided with that determined with a test object.

This approach allows estimating the MTF independently of the imaging modality.